\documentclass[aps,pre,preprintnumbers,floatfix,nofootinbib,12pt]{revtex4}
\usepackage{amsmath}
\usepackage{amssymb}
\usepackage{graphicx,color}
\usepackage{multirow}

\pdfoutput=1

\begin{document}

\title{Analysis of earlier times and flux of entropy on the majority
voter model with diffusion}
\author{Roberto da Silva$^{1}$, Mario J. de Oliveira$^{2}$, T\^ania Tom\'e$%
^{2}$, J. R. Drugowich de Fel\'icio$^{3}$}

\address{1 - Instituto de F\'{i}sica, Universidade Federal do Rio Grande do Sul, Av. Bento Gon\c{c}alves, 9500, 91501-970, Porto Alegre Rio Grande do Sul, Brazil\\
2 - Instituto de F\'{i}sica, Universidade de S\~{a}o Paulo, Rua do Mat\~{a}o, 1371, 055080-090, S\~{a}o Paulo, SP, Brazil\\
3 - Departamento de F\'{i}sica, Faculdade de Filosofia, Ci\^{e}ncias e Letras de Ribeir\~{a}o Preto, Universidade de S\~{a}o Paulo, Av. dos Bandeirantes 3900, 14040-901, Ribeir\~{a}o Preto, S\~{a}o Paulo, Brazil}

\begin{abstract}
We study the properties of nonequilibrium systems modelled as spin models
without defined Hamiltonian as the majority voter model. This model has
transition probabilities that do not satisfy the condition of detailed
balance. The lack of detailed balance leads to entropy production phenomena,
which are a hallmark of the irreversibility. By considering that voters can
diffuse on the lattice we analyze how the entropy production and how the
critical properties are affected by this diffusion. We also explore two
important aspects of the diffusion effects on the majority voter model by
studying entropy production and entropy flux via time-dependent and steady
state simulations. This study is completed by calculating some critical
exponents as function of the diffusion probability.
\end{abstract}

\maketitle

\section{Introduction}

\label{sec:introduction}

The study of nonequilibrium systems \cite{MarioBook,TTome2015} can be
divided in two situations: systems that remain out of thermodynamic
equilibrium even in the stationary regime, and systems that are out of
equilibrium because they had not reach equilibrium. In the latter case, the
systems are characterized by obeying, in the stationary regime, the detailed
balance condition \cite{MarioBook,Kampen1981}: 
\begin{equation}
\frac{w_{i}(\sigma )}{w_{i}(\sigma ^{i})}=\frac{P(\sigma ^{i})}{P(\sigma )}.
\label{Eq:detailed_balance}
\end{equation}%
From this condition we find that in equilibrium the system is described by
the Gibbs distribution. In the former case the detailed balance condition is
not satisfied. In this equation, $\sigma $ denotes the collection of the
variables $\sigma _{i}=\pm 1$, that is, $\sigma =(\sigma _{1},...,\sigma
_{N})$, and $\sigma ^{i}$ denotes the state obtained from $\sigma $ by
changing the sign of $\sigma _{i}$. In addition, $w_{i}(\sigma )$ is the
transition rate from $\sigma $ to $\sigma ^{i}$ and $P(\sigma )$ is the
stationary probability distribution. Here, $N$ is the number of sites in the
lattice. The equation that governs the time evolution of the probability
distribution $P(\sigma ,t)$ is the master equation \cite{MarioBook,TTome2015}
\begin{equation}
\frac{d}{dt}P(\sigma ,t)=\sum_{i}\left[ w_{i}(\sigma ^{i})P(\sigma
^{i},t)-w_{i}(\sigma )P(\sigma ,t)\right] .  \label{Eq:Master_equation}
\end{equation}

Irreversible systems are in a process of continuous entropy production even
in the steady steady. The main question to ask here is how to calculate the
entropy production. To answer this question, we start by writing the rate of
change of the Gibbs entropy 
\begin{equation}
S(t)=-\sum_{\sigma }P(\sigma ,t)\ln P(\sigma ,t),  \label{Eq:Shannon}
\end{equation}%
which is split into two parts 
\begin{equation}
\frac{dS}{dt}=\Pi -\Phi ,  \label{Eq:Prigogine}
\end{equation}%
where $\Pi $ is the entropy production rate due to irreversible processes
occurring inside the system and $\Phi $ is the flux of entropy from inside
to outside the system. The expression of the entropy production rate is \cite%
{MarioBook} 
\begin{equation}
\Pi =\frac{1}{2}\sum_{\sigma }\sum_{i}\left[ w_{i}(\sigma ^{i})P(\sigma
^{i})-w_{i}(\sigma )P(\sigma )\right] \ln \frac{w_{i}(\sigma ^{i})P(\sigma
^{i})}{w_{i}(\sigma )P(\sigma )}.  \label{Eq:Entropy_production}
\end{equation}%
From Eqs. (\ref{Eq:Master_equation}), (\ref{Eq:Shannon}), and (\ref%
{Eq:Prigogine}) we obtain the following expression for the flux of entropy 
\begin{equation}
\Phi =\frac{1}{2}\sum_{\sigma }\sum_{i}\left[ w_{i}(\sigma ^{i})P(\sigma
^{i})-w_{i}(\sigma )P(\sigma )\right] \ln \frac{w_{i}(\sigma ^{i})}{%
w_{i}(\sigma )}
\end{equation}%
or 
\begin{equation}
\Phi =\sum_{i}\sum_{\sigma }w_{i}(\sigma )P(\sigma ) \ln \frac{w_{i}(\sigma )%
}{w_{i}(\sigma ^{i})}.
\end{equation}%
In this form we see that the flux of entropy can be written as an average
over the probability distribution $P(\sigma ,t)$, that is, 
\begin{equation}
\Phi =\sum_{i}\left\langle w_{i}(\sigma ) \ln \frac{w_{i}(\sigma )}{%
w_{i}(\sigma ^{i})}\right\rangle ,  \label{Eq:Flux_entropy}
\end{equation}%
an expression that has been used to determine the flux of entropy by Monte
Carlo simulation \cite{Crochik2005}, and which will be used here.

We observe in Eq. (\ref{Eq:Entropy_production}), that $\Pi =0$ if the
condition given by Eq. (\ref{Eq:detailed_balance}), the detailed balance
condition, is satisfied, which happens when the system in the condition of
thermodynamic equilibrium. In a nonequilibrium steady state the entropy
production rate does not vanish although the rate of the entropy of the
system, $dS/dt$, vanishes. In this case $\Phi =\Pi \neq 0$ and the system is
in a continuous production of entropy. In this case, the rate of entropy
production rate $\Pi $ can be determined by Eq. (\ref{Eq:Flux_entropy})
because $\Pi =\Phi $. The quantity $\Phi $ given by the Eq. (\ref%
{Eq:Flux_entropy}) is calculated as an average over the stationary
distribution, which from numerical point of view can be estimated by an
average over Monte Carlo simulation obtained after a transient. We point out
that the relaxation of the flux of entropy in some way must be related to
the relaxation of magnetization and its moments.

The relaxation of spin systems toward the steady state has been considered
in the study of the time-dependent Monte Carlo simulations. This is carried
out by changing the average over Monte Carlo steps at steady state, by
considering thermodynamic quantities in the earlier times of the evolution,
taking the average over different time series that such system can follow,
considering not only the randomness effects of the evolution but also the
trace of the initial condition of the system.

The universality and scaling behavior even at the beginning of the time
evolution of such dynamical systems around the criticality, can be resumed
by the relation \cite{hinrichsen2000,janssen1989} 
\begin{equation}
m(t)\sim t^{-\beta /\nu z}f((p-p_{c})t^{1/\nu z},t^{d/z}L^{-d},m_{0}t^{\beta
/\nu z+\theta }),
\end{equation}%
which can be employed in equilibrium or nonequilibrium systems, considering
their respective characteristics, where $\beta $ and $\nu $ are static
exponents while $z$ and $\theta $ are dynamic exponents. In the present case 
\begin{equation}
m(t)=\frac{1}{N}\sum_{i}\langle \sigma _{i}\rangle
\end{equation}%
is the magnetization of the system understood as an average over a certain
number $N_{run}$ of runs.

For small initial magnetization $m_{0}<<1$, we expect to find a
characteristic initial slip $m(t)\sim t^{\theta }$, characterized by the
exponent $\theta $. On the other hand, if the initial magnetization is not
small, for instance $m_{0}=1$, then we expect $m(t)\sim t^{-\beta /\nu z}$.
However this power law corresponds actually to an intermediate regime that
occurs before the system reaches the stationary case. In a more complete
point of view we have 
\begin{equation}
m(t)=\left\{ 
\begin{array}{lll}
m_{0}\,t^{\theta }, &  & 0<t<m_{0}^{-z/x_{0}} \\ 
&  &  \\ 
t^{-\beta /\nu z}, &  & m_{0}^{-z/x_{0}}<t<t_{st}\ ,%
\end{array}%
\right.  \label{Eq:Initial_slip}
\end{equation}%
where $t_{st}$ is the time needed to reach the steady state. But an
important question concerns the behavior of the system related to the
flux of the entropy. Thus, in a manner similar to that employed in
short-time studies, we have adapted the expression (\ref{Eq:Flux_entropy})
by considering the average over different runs before the steady state is
reached, that is, we consider the following expression 
\begin{equation}
\phi (t,m_{0})=\frac{1}{N}\sum_{i}\left\langle w_{i}(\sigma ) \ln\frac{%
w_{i}(\sigma )}{w_{i}(\sigma ^{i})}\right\rangle ,  \label{Eq:flux_via_MC}
\end{equation}%
where the average is understood to be taken from several runs. In the
following we will see how this quantity is related to the short-time
behavior and to the short-time dynamics(STD).

Questions related to both the entropy production \cite{TTome2015,Crochik2005}
and short-time simulations \cite{JFFMendes} have already been answered for
the interesting case of the majority voter model (MVM), a model without
Hamiltonian in the kinetic-Ising universality class \cite{Oliveira1991}. In
this model the transition rate is given by 
\begin{equation}
w_{i}(\sigma ))=\frac{1}{2}\left[ 1-(2p-1)\sigma _{i}\,S\left( \sum_{\delta
}\sigma _{i+\delta }\right) \right] ,
\end{equation}%
where $\sigma _{i}=\pm 1$, and $S(x)=-1,0,1$, according to $x<0$, $x=0$, or $%
x>0$. This model can be interpreted as an Ising model in contact with two
heat baths at different temperatures, one at zero temperature and other one
at infinite temperature. Grinstein \textit{et al}. \cite{Grinstein} conjectured that
systems with up-down symmetry belong to the universality class of the
equilibrium Ising model for regular square lattices. This model also has a
interpretation within the social dynamics. A voter follows the majority with
probability $p$\ and changes his or her vote with probability $q=1-p$ (for a more
detailed social exploration of the model see for example Ref. \cite{Castellano}).
Whatever the interpretation, we believe that diffusive effects of the voters
might have important effects on the general behavior of the model and
possibly on its critical behavior of the model.

In this paper we propose to study the majority voter model with diffusion of
the voters focusing on the entropy production and flux of entropy by
employing a time-dependent Monte Carlo simulations in a two dimensional
lattice by means of the short-time dynamics. For that, we first propose to
use a recent refinement process based on the short-time dynamics \cite%
{silva2012} to determine the critical parameter $p_{c}$ as a function of the
mobility $\alpha $\ (probability that a voter randomly chosen in the lattice
changes its place with a nearest neighbor also randomly chosen). In
addition, we also analyze the effects of such mobility on the dynamic
exponents and on the entropy flux in the steady state by using these
parameters previously calculated via the refinement process.

Before carrying out the present study, we present some previous results as a
preparatory study for $\alpha =0$. We explore the transient of $\Phi (t)$
under the light of short-time dynamics. For this purpose, we revisit the
mean-field(MF) results obtained in Ref. \cite{Crochik2005} in order to adapt them to
the present context of short-time dynamics for the majority voter model and
thus to extract an expression for $\Phi (t)$ at criticality in the
mean-field regime.

\section{Mean field of the Majority voter model: Time-dependent entropy flux}

In Ref. \cite{Crochik2005} the authors have considered the time evolution of
magnetization: 
\begin{equation}
\frac{d\langle \sigma _{i}\rangle }{dt}=-2\langle \sigma _{i}w_{i}(\sigma
)\rangle ,
\end{equation}%
and observed that the sign function can be written as 
\begin{equation}
S(\sigma _{1}+\sigma _{2}+\sigma _{3}+\sigma _{4})=\frac{1}{8}(3-\sigma
_{1}\sigma _{2}\sigma _{3}\sigma _{4})(\sigma _{1}+\sigma _{2}+\sigma
_{3}+\sigma _{4}).
\end{equation}%
Using an approximation in which $\langle \sigma _{i}\sigma _{j}\sigma
_{k}\rangle =m^{3}$ where $m=\langle \sigma _{i}\rangle $, the following
equation for $m$ is obtained 
\begin{equation}
\frac{dm}{dt}=\left( \frac{3}{2}\gamma -1\right) m-\frac{\gamma }{2}m^{3},
\label{25}
\end{equation}%
where the parameter $\gamma $ is related to the parameter $p$ by $\gamma
=2p-1$. The solution of this equation is 
\begin{equation}
m=\frac{\left( 2-3\gamma \right) ^{1/2}}{\left( 2e^{(2-3\gamma
)(t+c)}-\gamma \right) ^{1/2}},  \label{Eq:mag_mean_field}
\end{equation}%
where the constant $c$ is related to the initial magnetization $m_{0}$ by 
\begin{equation}
c=\frac{1}{2-3\gamma }\ln \left( \frac{2-3\gamma +\gamma m_{0}^{2}}{%
2m_{0}^{2}}\right) .
\end{equation}%
For large times $m\sim e^{-t/\tau }$ where $\tau =2/(2-3\gamma )$ is the
time correlation length.

This exponential behavior turns into a power law at the critical point that
occurs when $\gamma =2/3$. At the critical point the solution of Eq. (\ref%
{25}) is 
\begin{equation}
m(t)=\frac{m_{0}}{\sqrt{1+2m_{0}^{2}\,t/3}},
\label{Eq:decay_magnetization_mean_field}
\end{equation}%
which leads to $m\sim t^{-1/2}$ for large times, independently of $m_{0}$.
Comparing with Eq. (\ref{Eq:Initial_slip}), we observe that no initial slip
is found in mean-field regime although Equation (\ref%
{Eq:decay_magnetization_mean_field}) shows a dependence on $m_{0}$. However
the stretched exponential behavior deviation from criticality corroborate
the results obtained via Monte Carlo simulations. However, the question is
whether this would bring behavior similar to the flux of entropy obtained
via Monte Carlo simulations and the answer is no, since the correct behavior
seems to be related to the initial slip of the magnetization as we will see
in the next section. But, it would be interesting to find a formula for the
time-dependence of the flux of entropy in this regime.

From the formula (\ref{Eq:Flux_entropy}) for the flux of entropy we obtain,
within the mean-field approximation, the following expression 
\begin{equation}
\phi =\frac{1}{16}\left[ -5\gamma +6(2-\gamma )m^{2}+(\gamma -4)m^{4}\right]
\ln \frac{q}{p}.  \label{Eq:flux_mean_field}
\end{equation}%
At criticality we are able to explicitly write down $\phi $ as function of $%
m_{0}$ and $t$ by taking into account the Eq. (\ref%
{Eq:decay_magnetization_mean_field}) 
\begin{equation}
\phi =\frac{5}{24}\ln 5+\frac{\ln 5}{2(m_{0}^{-2}+\frac{2}{3}t)^{2}}\left[ 
\frac{1}{3}-(m_{0}^{-2}+\frac{2}{3}t)\right] .  \label{Fig:mean_field_flux}
\end{equation}%
In the limit $t\rightarrow \infty $ we see that the flux of entropy
approaches a nonzero value $\phi _{\infty }=(5/24)\ln 5$. This value is
reached through the power law 
\begin{equation}
\phi -\phi _{\infty }\sim t^{-1}.  \label{Eq:power_law_for_flux}
\end{equation}

In the next section we will see how these results for the two-dimensional
voter model are modified when the Monte Carlo simulations are used. We also
show how the diffusion of the voters on the two-dimensional lattice affects
the entropy production at the steady state, and the short time properties.

\section{Monte Carlo simulations}

\label{sec:Results}

In this section we study numerically the majority voter model with diffusion
of the voters. We perform MC simulations on a square lattices
with periodic boundary conditions with $N=L^{2}$ sites. At each time step we
choose a site at random and we decide if it will be flip or not according to
signal of the sum of their neighbors. One MC step is defined as repeating
this procedure $N$ times. In each MC step, after the updating of the spins,
we perform the diffusion, that is, we choose $N$ random pairs of neighboring
sites and we swap their positions with probability $0\leq \alpha \leq 1$. We
perform the diffusion of the voters in full lattices, so dilution is not a
parameter here.

In steady state simulations we performed averages using $10^{6}$ MC steps to
calculate the flux $\phi $, after discarding $10^{3}$ MC steps. On the other
hand in short-time simulations, we perform $N_{run}=20 000$ different times
series to compute $m(t)$ and $\phi (t)$. Particularly for simulations
starting from ferromagnetic initial systems we have considered a more modest
number of runs $N_{run}=3000$ runs since in this situation, the initial
trace is not important and smaller fluctuations are observed. This reasoning
is often used in short-time studies. In the present paper, we obtain our
estimates for $L=128$ unless we explicitly study some lattice effects and
small sizes are explored in order to corroborate that this size is enough
for our purpose.

\begin{figure}[tbh]
\begin{center}
\includegraphics[width=0.9
\columnwidth]{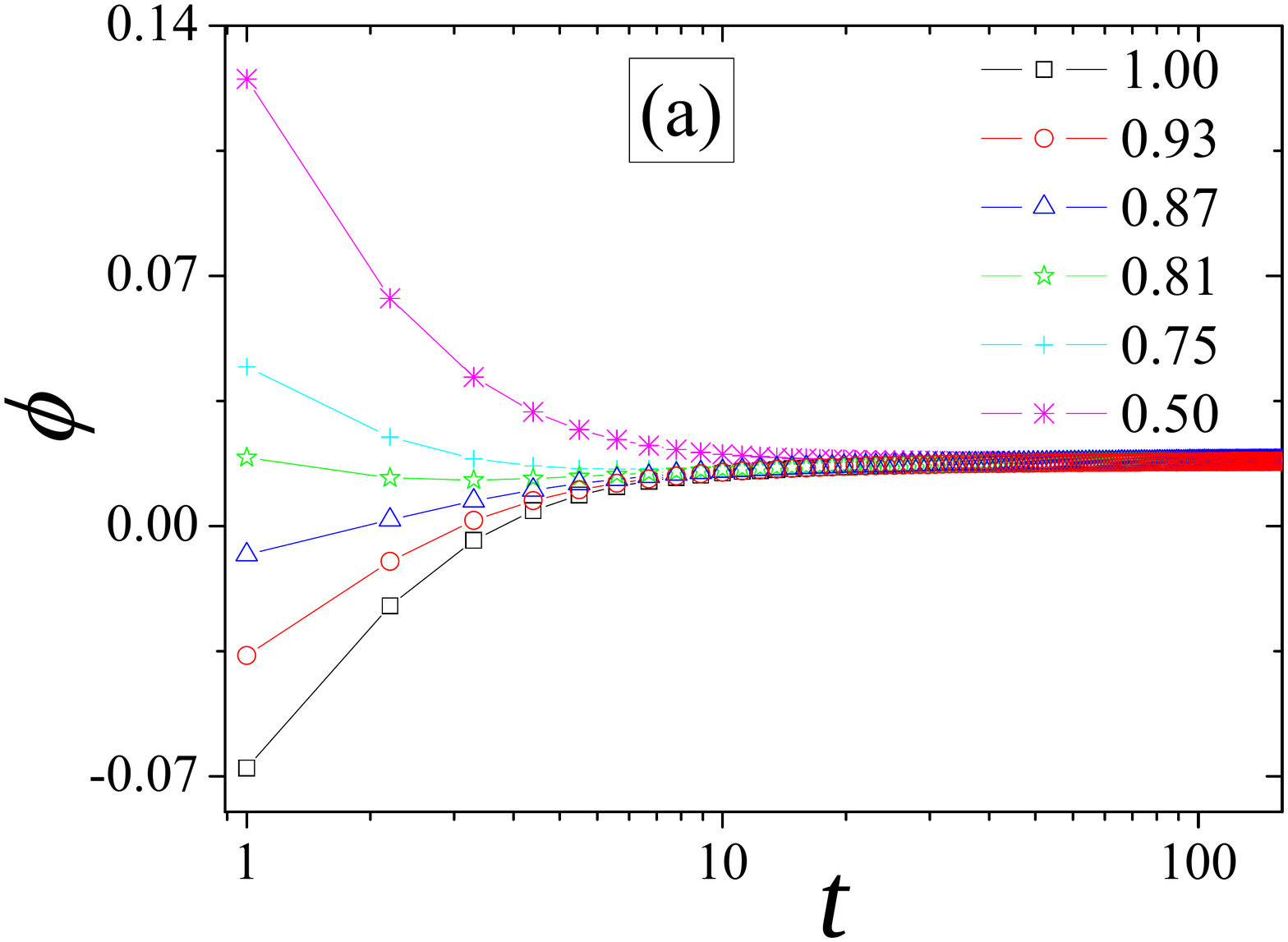} 
\includegraphics[width=0.9
\columnwidth]{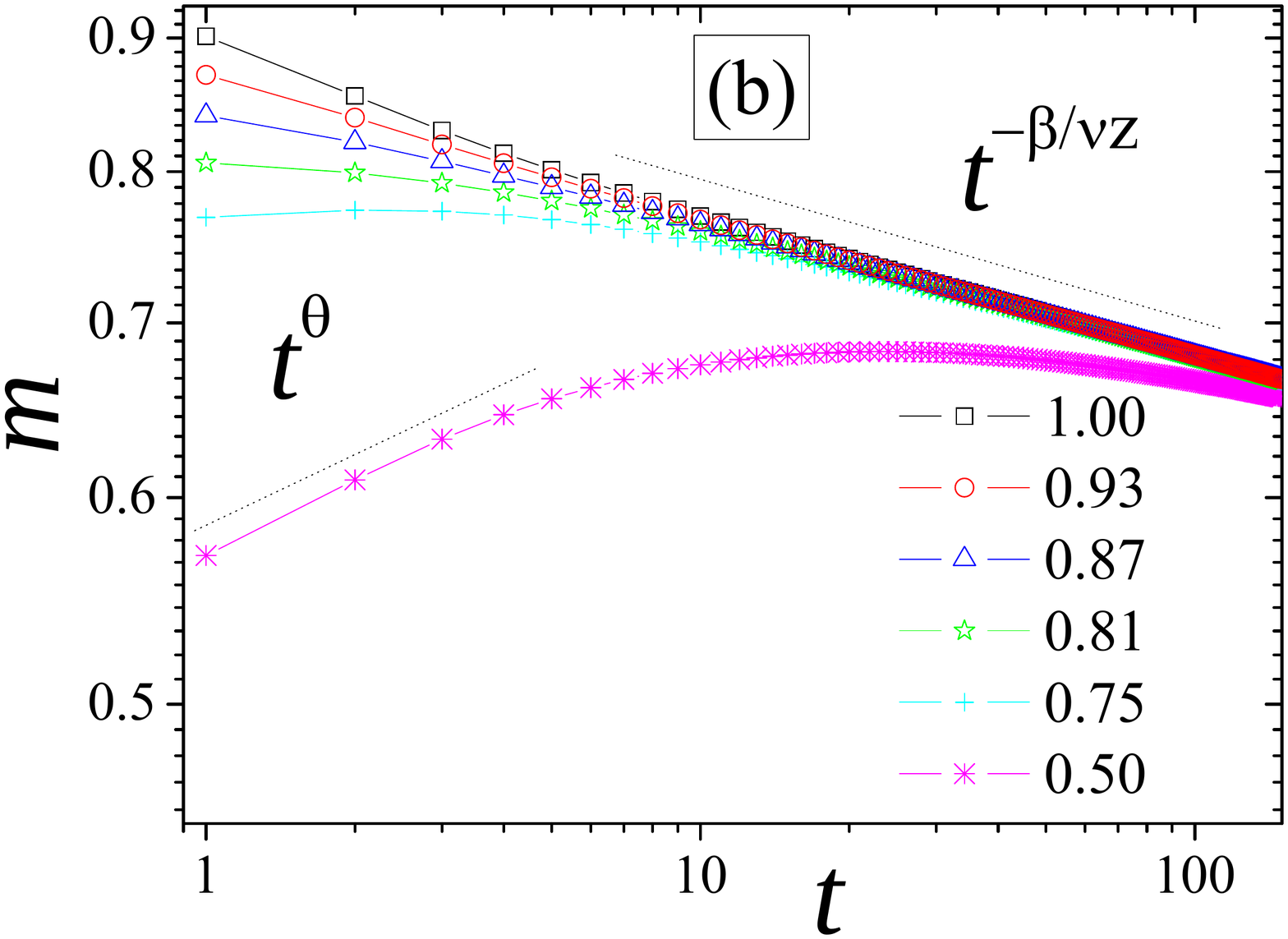}
\end{center}
\caption{(a): Time evolution of the entropy flux $\protect\phi $ obtained by
MC simulations for different initial magnetizations $m_{0}$ indicated. (b)
Time evolution of magnetization $m$ for different initial magnetizations $%
m_{0}$ indicated, obtained in the same simulations. }
\label{Fig:MC_simulations_alfa=0}
\end{figure}

In our first exploratory investigations, the voters are not subject to
diffusion ($\alpha =0$). Fig. \ref{Fig:MC_simulations_alfa=0}a shows the
behavior of $\phi (t)$ via Monte Cartlo simulations estimated according to
Eq. \ref{Eq:flux_via_MC} at the critical value $p_{c}=0.925$ known for the
model. Different initial magnetizations converge for the same value and the
flux may increase or decrease depending on its correlation level.

Figure \ref{Fig:MC_simulations_alfa=0}b shows the typical short-time behavior
expected for a spin model via MC simulations according to Eq. (\ref%
{Eq:Initial_slip}). Alternatively and only for a comparison, we can observe
that some differences can be observed in mean-field regime by using the
equations obtained in the previous section.

\begin{figure}[tbh]
\begin{center}
\includegraphics[width=1.0\columnwidth]{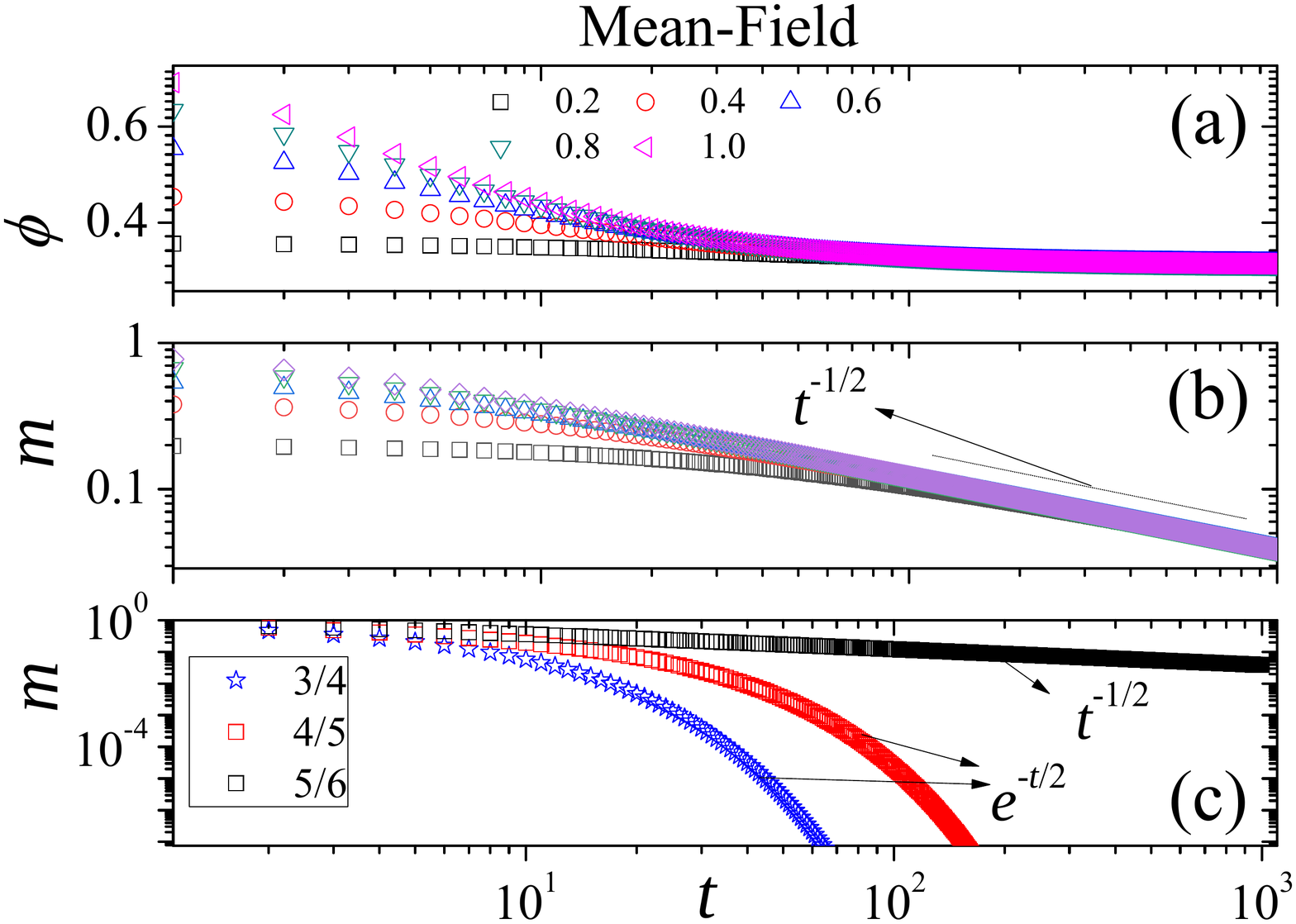}
\end{center}
\caption{Simple mean-field analysis of the system. (a): Time evolution of
the entropy flux $\protect\phi $ for different initial magnetizations $m_{0}$
indicated at criticality $p_{c}=5/6$. (b) The corresponding time evolution
of the magnetization $m$ also at criticality. (c) Deviation from $p=p_{c}$
of the magnetization $m$ for several values of $p$ indicated. The initial
condition is $m_{0}=1$. }
\label{Fig:Mean_field}
\end{figure}

First, Fig. \ref{Fig:Mean_field}(a) shows that the mean-field entropy flux $%
\phi $ according to Eq. \ref{Fig:mean_field_flux} independently of $m_{0}$
always decreases to reach the $\phi _{\infty }=(5/24)\ln 5\approx 0.33$
which is much bigger than the steady state value obtained for MC
simulations. Differently from the MC simulations, in Fig. \ref%
{Fig:Mean_field}(b) we do not observe a initial slip for magnetization and we
observe that $m(t)$ universally decays as $t^{-1/2}$. We believe that \
absence of initial slip in mean-field short-time behavior must affect the
differences on $\phi (t)$ via MC simulations and the MF approach. Finally, we
plot the stretched exponential deviation from this power law considering $%
p<p_{c}$, which is also expected in STD theory when studied by
time-dependent simulations, but we observe a more salient effect on the MF
regime.

\begin{figure}[tbh]
\begin{center}
\includegraphics[width=1.0\columnwidth]{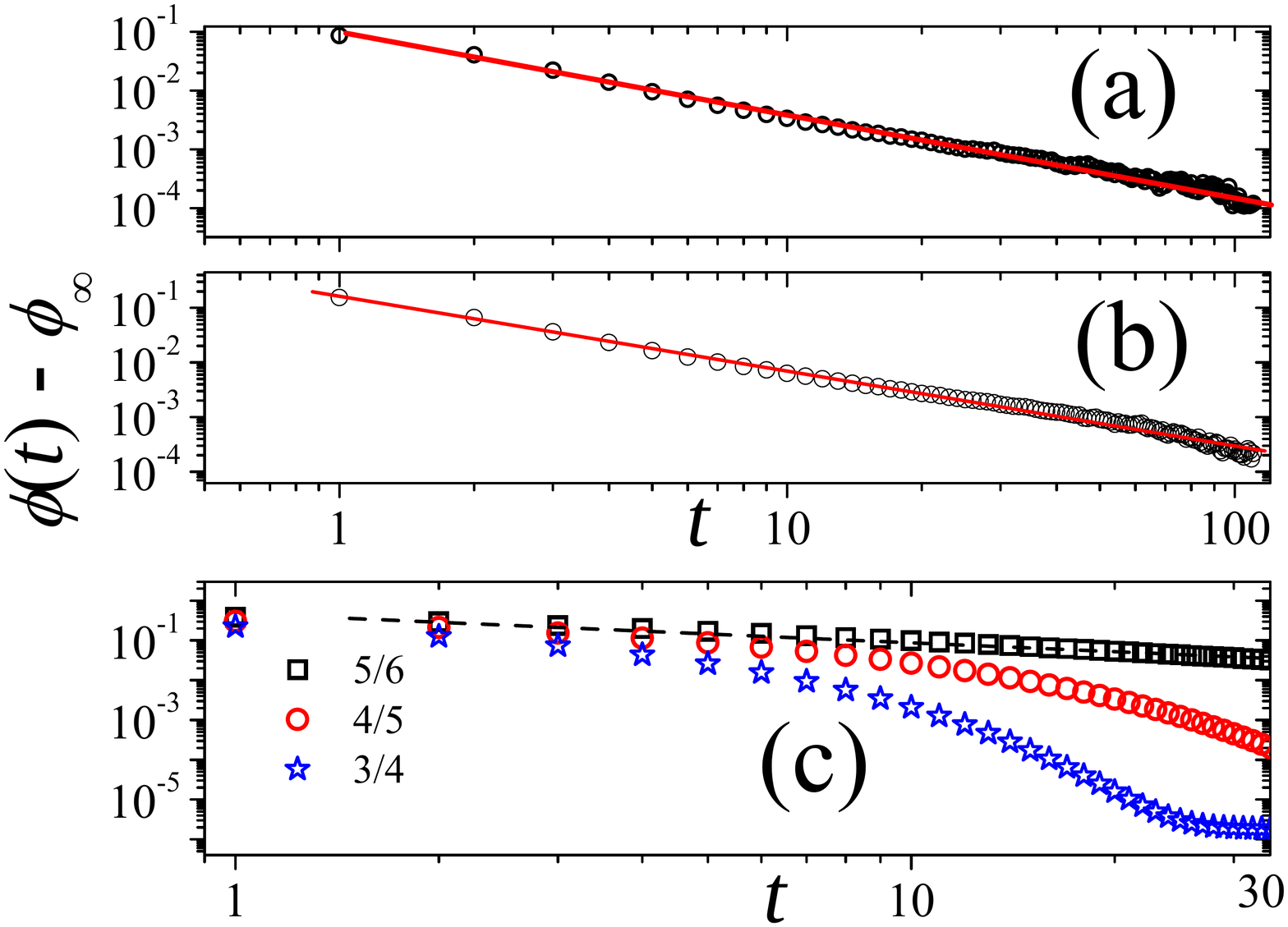}
\end{center}
\caption{Power law behavior of the quantity $\protect\phi -\protect\phi %
_{\infty }$ obtained MC simulations: (a) $m_{0}=1$, and (b) $m_{0}=1/2^{9}$,
and (c) the evolution of the same quantity obtained via mean-field for
different values of $p$: $p=p_{c}=5/6$, $p=4/5$, and $p=3/4$. }
\label{Fig:power_law_flux}
\end{figure}

It is also interesting to investigate the power law verified in mean-field
by Eq. (\ref{Eq:power_law_for_flux}). In this case we can study the quantity $%
\phi -\phi _{\infty }$ as function of $t$, expecting a power law behavior $%
\phi -\phi _{\infty }\sim t^{-\xi }$. Differently from mean-field regime, in
MC simulations we expect $\xi _{MC}\neq \xi _{MF}=1$. In Fig. \ref%
{Fig:power_law_flux} we present a study of the quantity $\phi -\phi _{\infty
}$ for two different situations: Fig. \ref{Fig:power_law_flux}a with
simulations starting from ferromagnetic inital state: $m_{0}=1$ and, Fig. %
\ref{Fig:power_law_flux}b considering an disordered initial state with a
very small initial magnetization, $m_{0}=1/2^{9}$. A stable power law can be
observed up to $t_{\max }\approx 110\ $MCsteps, and we obtained respectively 
$\xi =1.36(2)$ and $\xi =1.38(2)$ for $m_{0}=1$, and $m_{0}=1/2^{9}$ showing
that there is no numerical evidence about a dependence on the initial
condition of the system. We can observe that $\xi _{MC}>\xi _{MF}$, but in
both cases the power law behavior is verified. Fig. \ref{Fig:power_law_flux}%
c shows for a comparison, the time evolution of $\phi -\phi _{\infty }$\ via
mean field approximation calculated by substituting the result from the Eq: %
\ref{Eq:mag_mean_field} in the Eq. \ref{Eq:flux_mean_field}. For $%
p=p_{c}=5/6 $\ we can observe a power law behavior (Eq. \ref%
{Fig:mean_field_flux}) and an exponential deviation can be observed for $p\
\neq p_{c}$.

Thus, we explore the MC simulations for $\alpha >0$. First we used a
technique to localize and refine the critical parameters by considering a
refinement method proposed in 2012 by two of the authors of this paper \cite%
{silva2012}. This approach, which is based on the refinement of the
coefficient of determination of the order parameter, allows to locate phase
transitions of systems in a very simple way.

Considering that discarding a certain $N_{\min }$ MC steps is needed since
the universal behavior which we are looking for emerges only after a time
period sufficiently long to avoid the microscopic short-time behavior (see
for example some papers in several different models: with defined
Hamiltonian \cite{shorttime} and without defined Hamiltonian \cite%
{shorttime2}). We can define the coefficient of determination as in \cite%
{silva2012} (see Appendix, section \ref{sec:Appendix}) 
\begin{equation}
r=\frac{\sum\limits_{t=N_{\min }}^{N_{MC}}(\overline{\ln m}-a-b\ln t)^{2}}{%
\sum\limits_{t=N_{\min }}^{N_{MC}}(\overline{\ln m}-\ln m(t))^{2}},
\label{eq:coef_det}
\end{equation}%
where $N_{MC}$ is the total number of MC steps and generically 
\begin{equation*}
\overline{O}=\frac{1}{(N_{MC}-N_{\min }+1)}\sum\nolimits_{t=N_{\min
}}^{N_{MC}}O(t)\text{.}
\end{equation*}%
The value of $N_{min}$ depends on the details of the system in study and it
is related to the microscopic time scale, i.e., the time the system needs to
reach the universal behavior in short-time critical dynamics \cite%
{janssen1989}.

When the system is near the criticality ($p\approx p_{c}$), for $m_{0}=1$,
we expect that the order parameter follows a power law behavior $\overline{%
m(t)}\sim t^{-\beta /\nu z}\ $which, in $\log \times \log $ scale, yields a
linear behavior and $r$ approaches 1. In this case, we expect the slope $b$
to be a good estimate of $\beta /\nu z$. On the other hand, when the system
is out of criticality, there is no power law and $r\simeq 0$. Thus, we are
able to use the coefficient of determination $r$ to look for critical
points. Thus, the idea of the method is very simple: we just need to sweep
the parameter $p$ and find the point that possess $r\simeq 1$.

\begin{figure}[tbh]
\begin{center}
\includegraphics[width=1.0\columnwidth]{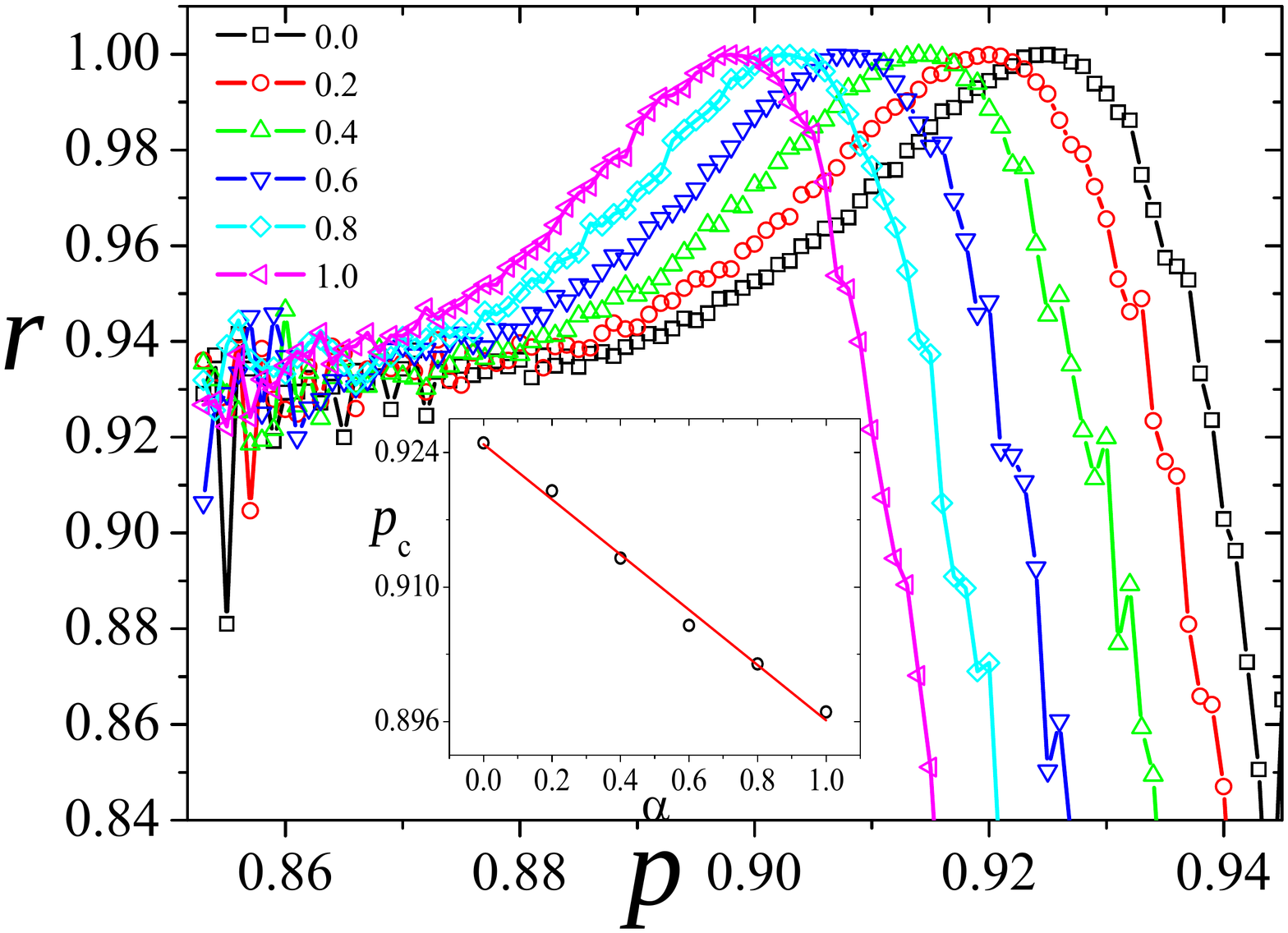}
\end{center}
\caption{Determination coefficient $r$ as function of $p$ for different
values of $\protect\alpha $ indicated. The peak of these curves (maximum
coefficient of determination correspond to the best $p_{c}$ obtained.}
\label{Fig:Determination}
\end{figure}

Thus, we performed simulations determining $p_{c}$ for each studied $\alpha $
studied (see Fig. \ref{Fig:Determination}), i.e., the best power
corresponding to $\overline{m(t)}\sim t^{-\frac{\beta }{\nu z}}$ which
occurs for ferromagnetic initial states, for example the case $\alpha =0$
can be observed in Fig. \ref{Fig:MC_simulations_alfa=0} (b).

\begin{figure}[tbh]
\begin{center}
\includegraphics[width=1.0\columnwidth]{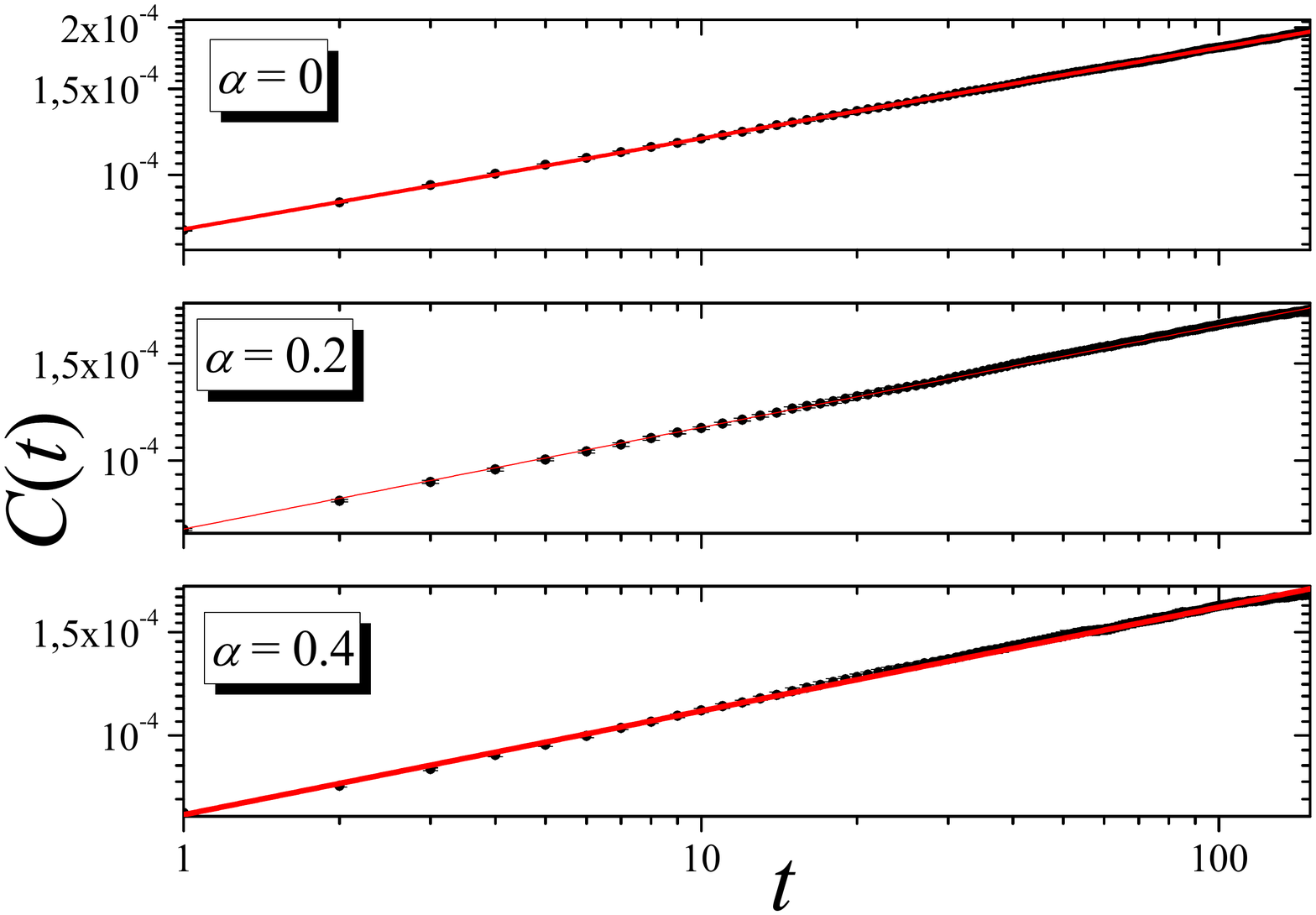}
\end{center}
\caption{Power law behavior of the correlation given by Eq. \protect\ref%
{Eq.Correlation} for three different diffusion levels. }
\label{Fig:correlation}
\end{figure}

The inset plot in Fig. \ref{Fig:Determination} shows the linear dependence
of $p_{c}$ as function of $\alpha $. We use a resolution of $\Delta
p=10^{-3} $. Thus, with these $p_{c}$ values in hands, we also determine $%
\theta =\theta (\alpha )$. For this task instead of studying the time
evolution of magnetization for small values of $m_{0}$, and thus performing
an extrapolation, we use the correlation of magnetization which also
presents a power law with this specific exponent \cite{TomeOliveira1998}%
: 
\begin{equation}
C(t)=\left\langle m(t)m(0)\right\rangle \sim t^{\theta },
\label{Eq.Correlation}
\end{equation}%
where in this case, $\left\langle m(0)\right\rangle \approx 0$. We can
observe in Fig. \ref{Fig:correlation} the power law from our MC simulations
for different mobility rates. In order to obtain the uncertainties on the
exponent $\theta $ we repeat the same simulations for $N_{b}=5$ different
beans, as well as the same procedure was used to obtain $\frac{\beta }{\nu z}
$ by using the relaxation of ferromagnetic initial states.

In table \ref{Table:Critical_values_and_exponents} we show the value of
exponents $\theta $ and $\beta /(\nu z)$ (last two columns) corresponding to
different values of diffusion $\alpha $. We also show the values of critical
parameters obtained by repeating the same optimization process for $N_{b}=5$
different seeds. We can observe that uncertainties leads to $\sigma
_{p}=O(10^{-4})$ which is better than resolution of optimization procedure ($%
\Delta =10^{-3}$).

\begin{table}[tbp] \centering%
\begin{tabular}{cccccccc}
\hline\hline
$\alpha $ & $p_{c}(1)$ & $p_{c}(2)$ & $p_{c}(3)$ & $p_{c}(4)$ & $p_{c}(5)$ & 
$\beta /(\nu z)$ & $\theta $ \\ \hline\hline
0.0 & 0.925 & 0.925 & 0.925 & 0.924 & 0.925 & 0.0541(13) & 0.185(6) \\ 
0.2 & 0.920 & 0.920 & 0.920 & 0.920 & 0.919 & 0.0642(15) & 0.176(6) \\ 
0.4 & 0.913 & 0.914 & 0.913 & 0.913 & 0.913 & 0.0779(17) & 0.160(2) \\ 
0.6 & 0.906 & 0.907 & 0.907 & 0.907 & 0.906 & 0.0932(36) & 0.151(6) \\ 
0.8 & 0.902 & 0.902 & 0.902 & 0.902 & 0.902 & 0.0996(10) & 0.14(1) \\ 
1.0 & 0.897 & 0.897 & 0.898 & 0.897 & 0.898 & 0.1108(29) & 0.12(1) \\ 
\hline\hline
\end{tabular}%
\caption{Critical parameters for 5 different repetitions of optimization and
the corresponding exponents for each value of diffusion rate}\label%
{Table:Critical_values_and_exponents}%
\end{table}%

We can observe that $\beta /(\nu z)$ increases while $\theta $ decreases as
the diffusion rate increases, showing that diffusion has a important role in
the phase transition of the model as observed for example in epidemic models
and surface reaction models (see for example \cite{dasilva2018-2015}). Just
for a comparison of the exponents with other results only for $\alpha =0$,
which is the only available estimates. In \cite{JFFMendes}, the authors
obtained $\theta =0.191(2)$ for MVM making extrapolation $m_{0}\rightarrow 0$
which in good agreement with our value. Both results also are in good
agreement with results for the Ising model obtained by Grassberger \cite%
{Grassberger} since both models are in the same universality class. It is
also important to mention that our result is close to the value of $\beta
/(\nu z)=0.0579(5)$, obtained by B. Zheng for the Ising model (see for
example first reference in \cite{shorttime}, pag. 1448).

By exploring some alternative results, we study the correlation of the
entropy flux $C_{F}(t)=\left\langle \phi (t)\phi (0)\right\rangle $ for
different diffusion rates. The results (see Fig. \ref%
{Fig:Correlation_entropy_flux}) show that correlation decay less quickly as
the diffusion increases.

\begin{figure}[tbh]
\begin{center}
\includegraphics[width=1.0\columnwidth]{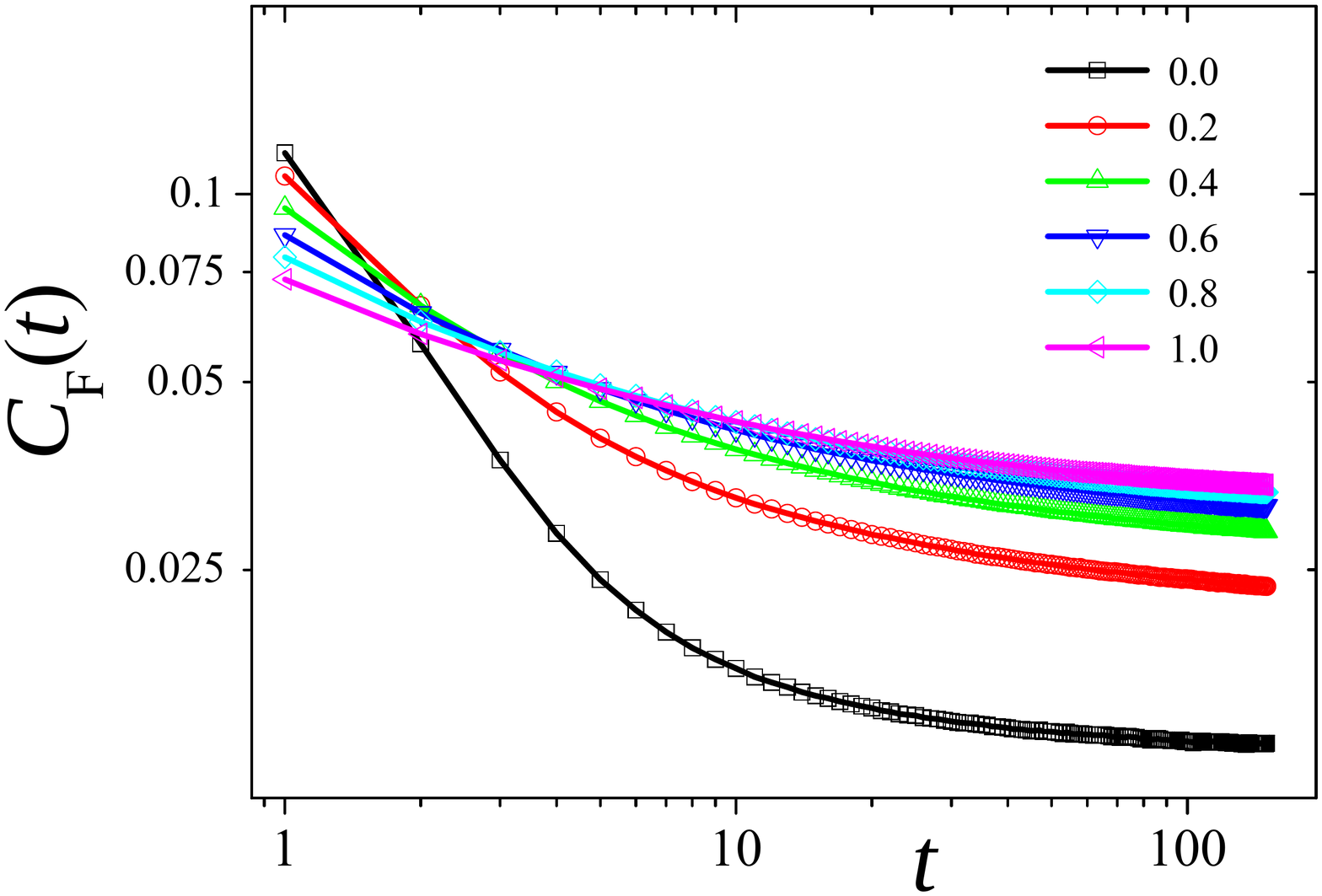}
\end{center}
\caption{Decay of the correlation of the entropy flux as function of time
for several values of $\protect\alpha $ indicated. }
\label{Fig:Correlation_entropy_flux}
\end{figure}

\begin{figure}[tbh]
\begin{center}
\includegraphics[width=0.9\columnwidth]{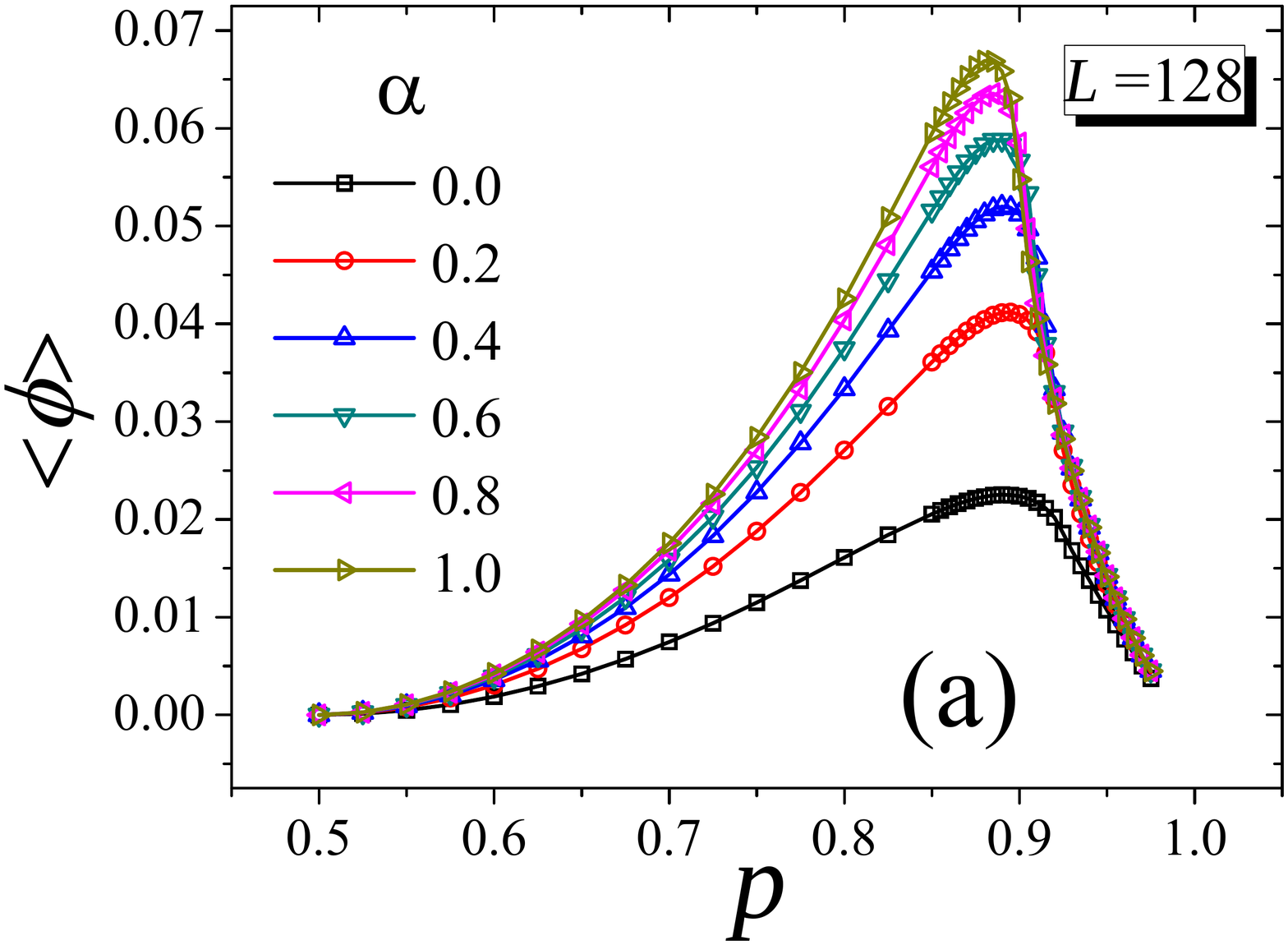} %
\includegraphics[width=0.9\columnwidth]{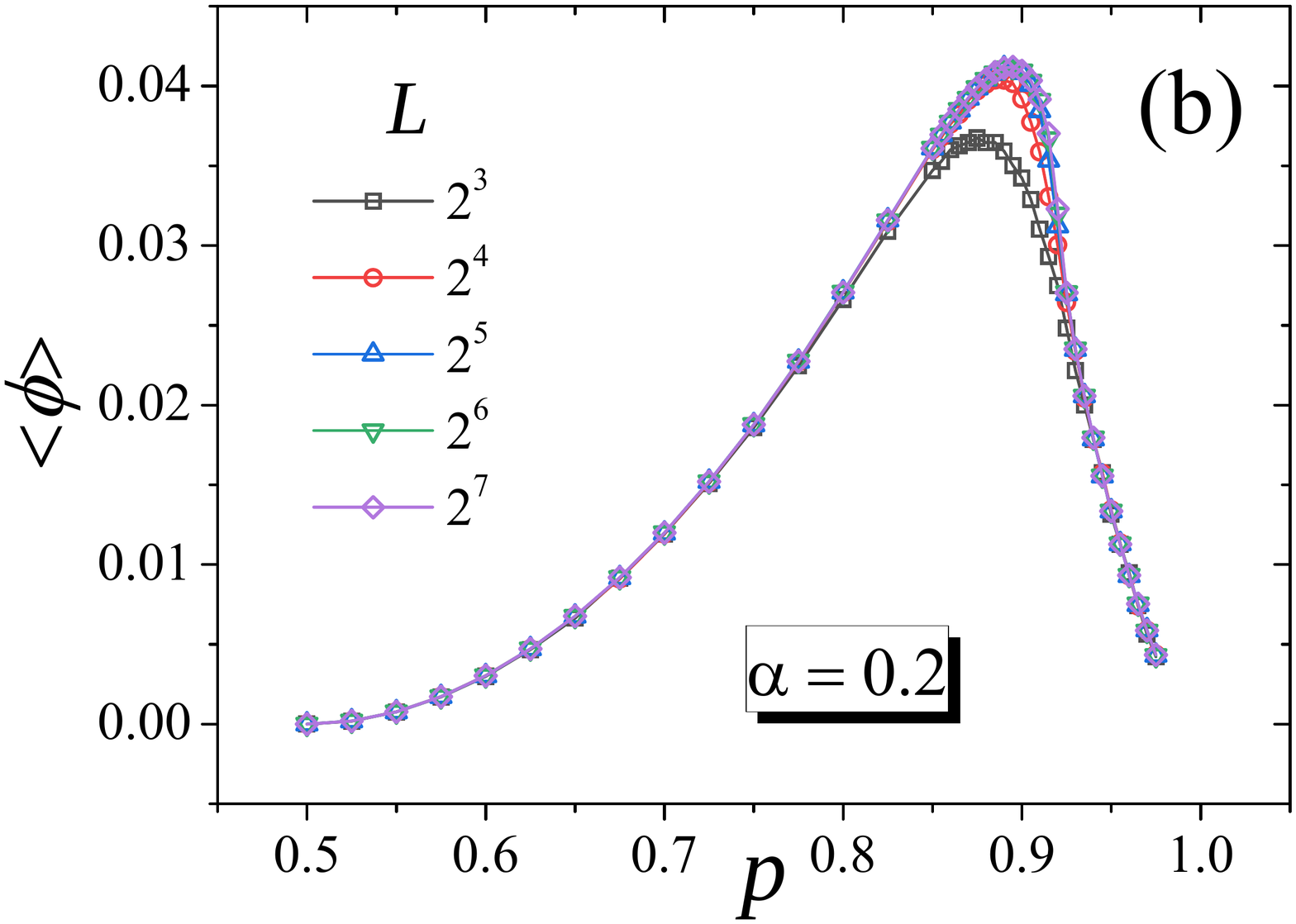}
\end{center}
\caption{(a) Entropy production of entropy calculated via flux at steady
state for different mobility rates for $L=128$. (b) Finite size effects
tested for a specific value of mobility: $\protect\alpha =0.2$. For $L\geq
32 $ we do not observe differences between the plots. }
\label{Fig:entropy_production}
\end{figure}

Since we have explored the time-dependent results, we finally explore the
effects of the diffusion on the entropy production which is equal to flux in
the steady state. Here we call attention of the readers that we look for $p$
and not $q$ as the authors in \cite{Crochik2005} studied. We can look that
our results for $\alpha =0$ recover that ones obtained by the authors (see
Fig. \ref{Fig:entropy_production} (a)) but, naturally, inverted.

This plot shows that entropy production (calculated as flux at steady state)
increases as the diffusion enlarges. Fig \ref{Fig:entropy_production} (b)
shows that for $L\geq 32$ we have no observed differences in the plots of
entropy production.

\section{Summary and Conclusions}

\label{sec:conclusions}

We have studied the diffusion effects on the Majority voter model exploring
time-dependent and time independent properties of the entropy flux. Our
results show that curve of entropy production enlarges as the diffusion rate
enlarge for any value of $p$. We also studied the effects of the diffusion
on the critical parameters of the model by using refinement procedure of
power laws in short time. Our results show that critical parameter depends
linearly on mobility rate. Similarly the dynamic exponents also depend on
mobility rate. Mean-field results are revisited and we make a empirical
comparison between short-time power law and entropy flux which still is a
preliminary study and deserves future investigation in other models.

\section{Appendix}

\label{sec:Appendix}

The coefficient of determination is a very simple concept used in linear
fits, or other fits. Thus, let us briefly explain such standard procedure in
the context of short-time MC simulations. When we perform least-square
linear fit to a given data set, we obtain a linear predictor $\widehat{y}%
_{t}=a+bx_{t}$. In addition, if we consider the unexplained variation given
by $\widetilde{\Delta }=\sum_{t=1}^{N}(y_{t}-\widehat{y}_{t})^{2}$ , and a
perfect fit is achieved when the curve is given by $y_{t}=a+bx_{t}$, and
therefore, $\widetilde{\Delta }=0$.

On the other hand, the explained variation $\Delta $ is given by the
difference between the average $\overline{y}=N^{-1}\sum_{t=1}^{N}y_{t}$, and
the prediction $\widehat{y}_{t}$, i.e., $\Delta =\sum_{t=1}^{N}(\widehat{y}%
_{t}-\overline{y})^{2}$. So, it is interesting to consider the total
variation, naturally defined as $\Delta _{total}=\sum_{t=1}^{N}(y_{t}-%
\overline{y})^{2}$. So, we can rewrite this last expression as $\Delta
_{total}=\sum_{t=1}^{N}(y_{t}-\widehat{y}_{t})^{2}+\sum_{t=1}^{N}(\widehat{y}%
_{t}-\overline{y})^{2}+\xi $, where $\xi =2\sum_{t=1}^{N}(y_{t}-\widehat{y}%
_{t})(\widehat{y}_{t}-\overline{y})$. However we can easily show that $\xi
=0 $, since

\begin{equation}
\begin{array}{lll}
\sum_{t=1}^{N}(y_{t}-c_{a}-c_{b}x_{t})(c_{a}+c_{b}x_{t}-\overline{y}) & = & 
c_{b}\sum_{t=1}^{N}x_{t}(y_{t}-c_{a}-c_{b}x_{t}) \\ 
&  & +\ (c_{a}-\overline{y})\sum_{t=1}^{N}x_{t}(y_{t}-c_{a}-c_{b}x_{t}) \\ 
&  &  \\ 
& = & -\ \frac{c_{b}}{2}\frac{\partial }{\partial c_{b}}%
\sum_{t=1}^{N}(y_{t}-c_{a}-c_{b}x_{t})^{2} \\ 
&  & -\ \frac{(c_{a}-\overline{y})}{2}\frac{\partial }{\partial c_{a}}%
\sum_{t=1}^{N}(y_{t}-c_{a}-c_{b}x_{t}),%
\end{array}%
\end{equation}%
and the last two sums vanish by definition when take the least squares
values $(c_{a},c_{b})=(a,b)$. Therefore, the total variation can be simply
defined as 
\begin{equation}
\Delta _{total}=\widetilde{\Delta }+\Delta \ ,
\end{equation}%
and the better the fit, the smaller the $\widetilde{\Delta }$. So, in an
ideal situation $\widetilde{\Delta }=0$, and thus the ratio 
\begin{equation}
r=\frac{\Delta }{\Delta _{total}}=1\ ,
\end{equation}%
i.e., the variation comes only from the explained sources.

So adapting this method to time-dependent MC simulations, if we consider
that $y_{t}=\ln m(t+N_{\min })$, $x_{t}=\ln (t+N_{\min })$, where $N_{\min }$
is the number of MC steps discarded at the beginning of the simulation (the
first steps), we can obtain the Eq. \ref{eq:coef_det}.

\section*{Acknowledgments}

R. da Silva thanks CNPq for financial support under grant numbers
311236/2018-9, and 424052/2018-0. This research was partially carried out
using the computational resources from the Cluster-Slurm, IF-UFRGS.

\end{document}